%% file: 0_main_arxiv.tex
\PassOptionsToPackage{table}{xcolor}
\documentclass[sigplan,10pt,screen]{acmart}
\acmYear{2026}
\acmISBN{} 
\acmDOI{} 
\startPage{1}

\setcopyright{none}

\usepackage{booktabs}   
\usepackage{subcaption} 
\usepackage{tikz}
\usepackage{amsmath}
\usepackage{subcaption}
\usepackage{multirow}
\usepackage{pifont}
\usepackage{listings}
\usepackage[linesnumbered, noend, ruled, vlined]{algorithm2e}
\usepackage{xcolor}
\usepackage{subcaption}
\usepackage{multirow}
\usepackage{array}
\usepackage{graphicx}
\usepackage{color}

\usepackage[table]{xcolor}
\definecolor{ZpfGreen}{RGB}{0,100,0}
\definecolor{ZpfRed}{RGB}{255,0,102}

\usepackage{filecontents}

\newcommand{\stitle}[1]{\vspace*{0.4em}\noindent{\bf #1.\/}}
\newcommand{\stitlestart}[1]{\noindent{\bf #1.\/}}

\newcommand{\squishlist}{
	\begin{list}{$\bullet$}
		{ \setlength{\itemsep}{1pt}
			\setlength{\parsep}{1pt}
			\setlength{\topsep}{2.5pt}
			\setlength{\partopsep}{0.5pt}
			\setlength{\leftmargin}{1em}
			\setlength{\labelwidth}{1em}
			\setlength{\labelsep}{0.6em}
		}
	}
	\newcommand{\squishend}{
	\end{list}
}

\newcommand{\name}{{SparseServe}}

\begin{document}

\title[PPoPP Paper]{\name{}: Unlocking Parallelism for Dynamic Sparse Attention in Long-Context LLM Serving}         


\author{Qihui Zhou}
\affiliation{
  \institution{The Chinese University of Hong Kong}
}
\email{qhzhou@cse.cuhk.edu.hk}   

\author{Peiqi Yin}
\affiliation{
  \institution{The Chinese University of Hong Kong}
}
\email{pqyin22@cse.cuhk.edu.hk}  

\author{Pengfei Zuo}
\authornote{Pengfei Zuo is the corresponding author.}
\affiliation{
  \institution{Huawei Cloud}
}
\email{pengfei.zuo@huawei.com}  

\author{James Cheng}
\affiliation{
  \institution{The Chinese University of Hong Kong}
}
\email{jcheng@cse.cuhk.edu.hk}  
\fancyhead{}  
\renewcommand\footnotetextcopyrightpermission[1]{} 

\begin{abstract}
Serving long-context LLMs is costly because attention computation grows linearly with context length. Dynamic sparse attention algorithms (DSAs) mitigate this by attending only to the key-value (KV) cache of critical tokens. However, with DSAs, the main performance bottleneck shifts from HBM bandwidth to HBM capacity: KV caches for unselected tokens must remain in HBM for low-latency decoding, constraining parallel batch size and stalling further throughput gains. Offloading these underutilized KV caches to DRAM could free HBM capacity, allowing larger parallel batch sizes. Yet, achieving such hierarchical HBM-DRAM storage raises new challenges, including fragmented KV cache access, HBM cache contention, and high HBM demands of hybrid batching, that remain unresolved in prior work.

This paper proposes SparseServe, an LLM serving system that unlocks the parallel potential of DSAs through efficient hierarchical HBM-DRAM management. SparseServe introduces three key innovations to address the challenges mentioned above: (1) fragmentation-aware KV cache transfer, which accelerates HBM-DRAM data movement through GPU-direct loading (FlashH2D) and CPU-assisted saving (FlashD2H); (2) working-set-aware batch size control that adjusts batch sizes based on real-time working set estimation to minimize HBM cache thrashing; (3) layer-segmented prefill that bounds HBM use during prefill to a single layer, enabling efficient execution even for long prompts. Extensive experimental results demonstrate that SparseServe achieves up to 9.26$\times$ lower mean time-to-first-token (TTFT) latency and up to 3.14$\times$ higher token generation throughput compared to state-of-the-art LLM serving systems.


\end{abstract}

\maketitle

\input{1_introduction}
\input{2_background}

\input{3_design}

\input{4_implementation}
\input{5_evaluation}
\input{6_related_work}
\input{7_conclusion}

\bibliography{sample-base}

\input{appendices}

\end{document}

%% file: 1_introduction.tex
\section{Introduction}\label{sec:introduction}

The rapid advancement of large language models (LLMs) has reshaped our daily lives. As the demand for long-context applications such as sophisticated reasoning~\cite{cot, tot, got} and document analysis~\cite{summ1, summ2} continues to grow, the ability of LLMs to process long sequences has become increasingly critical. To meet this need, recent models have expanded their context windows to over one million tokens~\cite{gemini,llama-3-8B-Gradient,lwm}.

However, serving long-context LLMs at scale incurs extremely high inference costs because attention computation grows linearly with sequence length. At each decoding step, all key-value (KV) cache entries are fetched from HBM to compute units. Since the KV cache itself expands linearly with sequence length, attention computation cost increases proportionally and becomes bounded by HBM bandwidth.


To reduce the inference cost of long-context LLMs, dynamic sparse attention algorithms (DSAs)~\cite{infllm, quest, arkvale, Yuan2025NativeSA, deepseekai2024deepseekv32} compress the KV cache during decoding, thereby reducing HBM bandwidth demands. DSAs exploit the observation that only a small set of critical tokens largely determines the output token, and that token criticality varies across query tokens. In self-attention, these critical tokens exhibit significantly higher attention scores ($Q^{T}K$) than others, allowing accurate approximation of attention using only their KV cache.
DSAs implement this through a select-then-compute manner: they partition the KV cache into blocks of consecutive tokens, maintain compact metadata for each block, and for every query token, estimate block criticality from its metadata to select the top-$k$ KV blocks for approximate attention. 

However, with DSAs, the main performance bottleneck shifts from HBM bandwidth to HBM capacity. To ensure low-latency decoding under uncertain token selection, all KV blocks must reside in HBM, even though only a small subset is accessed for each decoding step. This leads to poor HBM capacity utilization and limits the parallel potential of DSAs to scale throughput by increasing batch sizes. 

Hierarchical memory management—where offloading all KV blocks to host memory (DRAM) while dynamically fetching only critical blocks into HBM for attention computation—offers a promising solution for reducing HBM pressure~\cite{infllm, tokenselect}. However, naive offloading can introduce significant decoding latency and degrade system throughput due to the limited DRAM access bandwidth. Enabling efficient hierarchical memory management for DSAs faces three key challenges that prior work has not addressed.



\textbf{\textit{1) Fragmented KV cache transfers reduce effective DRAM access bandwidth.}}
The physical bandwidth from device (GPU) to host memory (DRAM) is limited. For an NVIDIA A100 40GB GPU, the D2H bandwidth via PCIe Gen4 is only ~32 GB/s, compared to ~1.6 TB/s for HBM. This bottleneck is further exacerbated by fragmented KV cache accesses of DSA. Specifically, existing DSAs~\cite{quest, arkvale} typically set the KV block size to 32 tokens and store the KV blocks of each attention head separately, which results in only 16 KB per block for popular long-context model like LWM-7B~\cite{lwm}. Fetching such small blocks via standard \texttt{cudaMemcpy} achieves an effective bandwidth of less than 4 GB/s, far below the PCIe peak and orders of magnitude lower than HBM.

\begin{figure}[!t]
	\centering
        \setlength{\abovecaptionskip}{0.1cm}
	\includegraphics[width=0.8\columnwidth]{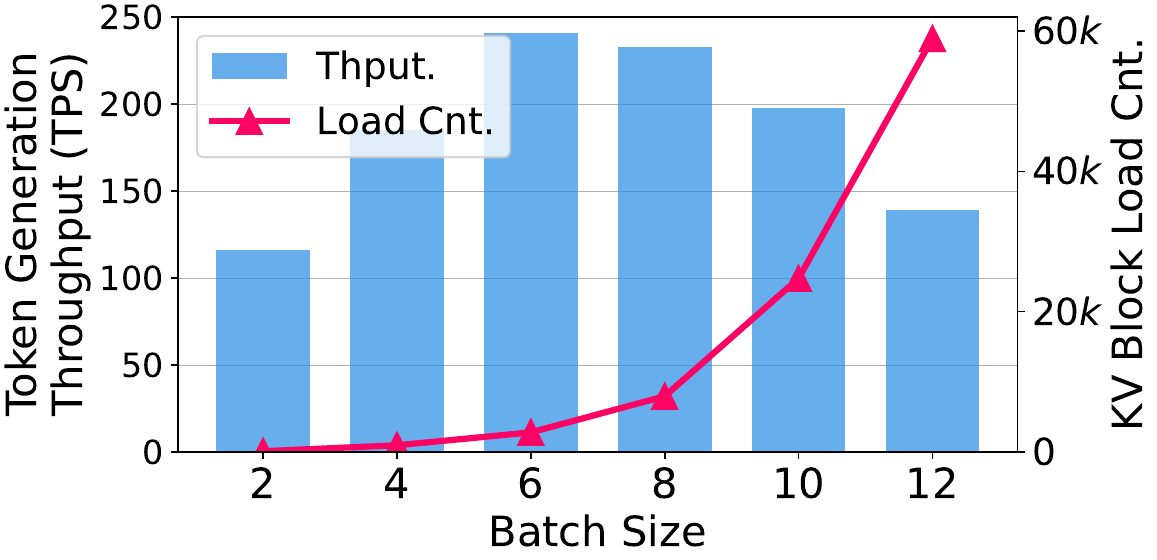}
        \caption{Token generation throughput and average number of KV blocks loaded per iteration under varying batch sizes.}
        \label{fig:bk:batch_size}
	\vspace{-3mm}
\end{figure}

\textbf{\textit{2) Efficient runtime control of parallel batch sizes.}}
For DSAs, small batch sizes underutilize GPU resources, limiting throughput. However, larger batch sizes do not always improve throughput. This is because large batch sizes may exacerbate HBM cache contention, leading to frequent KV cache evictions and increased data transfers from DRAM. 
Balancing batch size dynamically at runtime is therefore critical to maximize efficiency while avoiding HBM bottlenecks.
We measure throughput and the average number of KV blocks loaded per iteration across varying batch sizes, as shown in Figure~\ref{fig:bk:batch_size}. The throughput initially increases by 2.07$\times$ when the batch size grows from 2 to 6. However, further increasing in batch size leads to a notable drop. For instance, expanding the batch size from 6 to 12 reduces throughput by 1.73$\times$. The decline is caused by the excessive KV cache loading associated with larger batch sizes. Specifically, the average number of KV blocks loaded per iteration increases by 21.36$\times$, when increasing the batch size from 6 to 12.


\textbf{\textit{3) High HBM requirement of chunked prefill.}}
To enhance GPU utilization, modern LLM serving systems widely adopt hybrid batching, combining compute-intensive prefill and memory-intensive decoding requests within the same batch. To prevent generation stalls, long prompts are divided into smaller chunks and processed over multiple batches, which is known as chunked prefill~\cite{sarathi}. While chunked prefill effectively reduces the computation per iteration, it cannot reduce the HBM consumption of request prefilling, since executing each prefill chunk requires the KV cache generated by all preceding chunks. As a result, requests with long input prompts may experience head-of-line blocking, as they must wait for currently running requests to complete and release sufficient HBM for execution.


To address these challenges, this paper introduces \name{}, an LLM serving system with efficient hierarchical HBM-DRAM management designed for efficient deployment of DSAs in long-context inference. To unlock the full parallel potential of DSAs, \name{} incorporates three key system innovations to overcome the deployment bottlenecks.

Firstly, to accelerate fragmented KV cache loading and saving, \name{} introduces \textit{FlashH2D} and \textit{FlashD2H}, two fragmentation-aware transfer engines that optimize KV cache movement between GPU HBM and host DRAM for DSAs. FlashH2D exploits the unified virtual addressing (UVA) capabilities of modern GPUs to perform GPU-direct loading, fusing multiple small KV block reads into a single kernel execution. FlashD2H adopts CPU-assisted saving, which first transfers the contiguous but unorganized KV cache into a DRAM buffer through a single \texttt{cudaMemcpy} and then utilizes CPU threads to asynchronously scatter the data into the corresponding KV blocks.

Secondly, to determine the optimal request batch sizes, \name{} employs a \emph{working-set-aware batch size control strategy} that dynamically adjusts batch sizes at runtime to avoid HBM cache thrashing. By exploiting the temporal locality of block selection, the working set of frequently accessed KV blocks can be estimated from previous iterations. \name{} ensures that the aggregated working set of scheduled requests stays within HBM capacity, reducing excessive KV cache loading resulting from cache contention.

Finally, to limit the HBM footprint during prefill, \name{} introduces \emph{layer-segmented prefill}, a new mechanism that conducts prefill layer by layer. This method evicts KV blocks of preceding layers to DRAM as subsequent layers are processed, thereby bounding the KV cache footprint of prefill to a single layer. To prevent generation stalls, each layer is further divided into multiple segments with each segment processed in a separate batch, significantly reducing batch running time.

We implement \name{} based on vLLM~\cite{vllm} and evaluate its performance using two popular long-context LLMs on the LongBench~\cite{longbench} dataset. Experimental results show that, compared to vanilla vLLM, \name{} reduces the mean time-to-first-token (TTFT) latency by up to 9.26$\times$ under the same request rates and improves token generation throughput by up to 3.14$\times$. The experimental results also indicate that the designs of \name{} are effective in improving the maximum request throughput under the service level objective (SLO) requirements. To summarize, this paper makes the following contributions:

\squishlist

    \item We identify the parallel potential of DSAs and highlight three fundamental challenges that limit its realization.

    \item We propose \name{}, a long-context LLM serving system that unlocks this parallelism through efficient hierarchical HBM-DRAM management, featuring fragmentation-aware KV cache transfer, working-set-aware batch size control, and layer-segmented prefill.

    \item We implement \name{} atop vLLM~\cite{vllm} and extensively evaluate it, demonstrating substantial performance improvements over vanilla vLLM and vLLM enhanced with state-of-the-art DSA.

\squishend

%% file: 2_background.tex
\section{Background and Motivation}


\subsection{Generative LLM Inference Basics}\label{subsec:llm-basics}
\stitlestart{Transformer}
The transformer has emerged as the standard model architecture for modern LLMs~\cite{gpt4,llama3.1}. These LLMs are typically composed of a chain of transformer layers, each containing two basic modules: self-attention and feed-forward network (FFN). During inference, the input query token list $X=[x_{1},x_{2},...x_{N_{q}}]$ for each layer is first multiplied by three weight matrices $W_{q}$, $W_{k}$, and $W_{v}$ to generate query ($Q \in R^{N_{q} \times d}$), key ($K \in R^{N_{kv} \times d}$), and value ($V \in R^{N_{kv} \times d}$) matrices, where $N_{q}$ is the number of query tokens, $N_{kv}$ is the number of KVs, and $d$ is the hidden dimension. The self-attention is then conducted using  $Q$, $K$, and $V$:
\begin{equation}
        S = {Q \cdot K^{T}} / {\sqrt{d}},\quad
        P = softmax(S),\quad
        O = P \cdot V
        \notag
\end{equation}
Here, $P$ represents the \emph{attention weights}, with $P_{i, j}$ indicating the \emph{importance} of $K_j$ to query token $i$. The attention output $O$ is then fed into the FFN module. The FFN output serves as the input of the next transformer layer.

\stitle{Auto-regressive generation}
Generative LLM inference proceeds in two phases: prefill and decoding. The prefill phase processes the input prompt in parallel to produce the first output token, with the prefill latency is measured by time-to-first-token (TTFT). The decoding phase then generates tokens auto-regressively, where each new token becomes the input for the next iteration. The latency for each generated token is measured by time-between-token (TBT).

\stitle{KV cache}
During the generation of new tokens, the key and value vectors of all previous tokens are required for the self-attention computation. As a result, they are cached in HBM to avoid repeated computation, referred to as KV cache. Since the KV cache grows with the decoding process and the number of decoding steps is unknown, existing systems generally employ PagedAttention~\cite{vllm} to divide HBM into fixed-size blocks and store the KV cache in multiple discontinuous blocks to avoid HBM fragmentation.


\stitle{Hybrid batching and chunked prefill}
During the prefill phase, all prompt tokens are processed in parallel in a single iteration, enabling efficient utilization of GPU compute resources. In contrast, the decoding phase involves a full forward pass of the LLM model over a single token generated in the previous iteration. This leads to low compute utilization and makes decode memory-bound. To improve GPU utilization, hybrid batching~\cite{sarathi, sarathiserve, nanoflow} is proposed to combine the prefill and decoding phases of different requests into the same batch. However, due to the processing of a large number of tokens, the latency of a prefill iteration is significantly higher than that of a decoding iteration, leading to high TBT. To address this issue, each input prompt for prefill is divided into multiple chunks, i.e., chunked prefill~\cite{sarathi}. These chunks are scheduled per iteration with ongoing decoding requests, effectively reducing the TBT.

\begin{figure}[!t]
	\centering
	\includegraphics[width=0.9\columnwidth]   {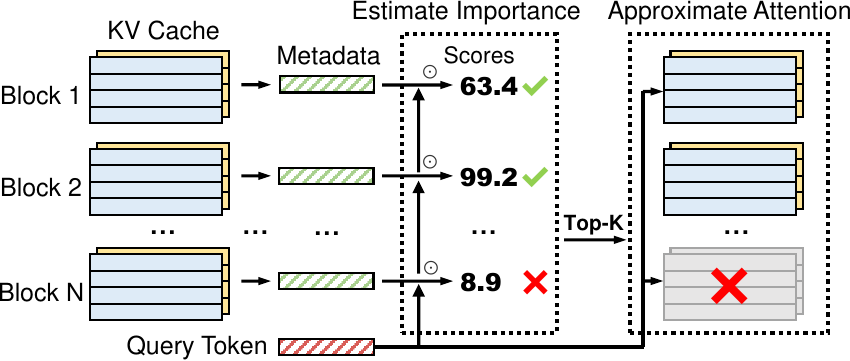}
	\caption{Workflow of dynamic sparse attention.}
	\label{fig:bk:dsaes}
	\vspace{-4mm}
\end{figure}


\subsection{Dynamic Sparse Attention}\label{subsec:dsaes} 
Due to the auto-regressive nature of LLMs, generating each token necessitates loading the entire KV cache from GPU HBM to on-chip SRAM, which results in significant time and space overheads for long-context LLM serving. 

Recent works~\cite{h2o, streamingllm} have observed that the attention computation is highly sparse, with only a small portion of tokens contributing to the majority of attention weights. Based on this observation, dynamic sparse attention algorithms (DSAs)~\cite{arkvale, infllm, quest} have been proposed. Figure~\ref{fig:bk:dsaes} illustrates the general workflow of existing DSAs, where a small portion of critical KV cache for each query token are dynamically selected for attention computation. Since not all KV cache is required for attention computation, DSAs allow KV cache to be offloaded to DRAM and load only the selected KV blocks into GPUs each time to reduce HBM consumption. 

To speed up the selection process, inspired by the block-level memory allocation of KV cache in PagedAttention~\cite{vllm}, DSAs divide  KV cache into blocks and select  KV cache at the block level. For each KV block, DSAs construct metadata vectors to represent the tokens within it. Different DSAs propose various metadata construction methods, ranging from simply calculating the mean values of the token keys~\cite{infllm} to finding the bounding cuboid of the token keys~\cite{arkvale}. Regardless of the methods, the size of the metadata is much smaller than the KV block. To estimate the importance of KV blocks to each query token, DSAs compute dot products between the metadata vectors and the query token to obtain approximate attention scores for all KV blocks. DSAs then select the top-$k$ most critical KV blocks to perform attention.

\subsection{Motivation and Challenge}

With DSAs, the main performance bottleneck shifts from HBM bandwidth to HBM capacity. To guarantee low-latency decoding under uncertain token selection, all KV blocks must be kept in GPU HBM, even though only a small subset is accessed at each decoding step. This results in inefficient HBM utilization and constrains the ability of DSAs to scale throughput by increasing batch sizes.

A promising solution is hierarchical memory management, which offloads all KV blocks to host DRAM while dynamically fetching only the critical blocks into HBM for attention computation~\cite{infllm, tokenselect}. This approach alleviates HBM capacity pressure and opens up more room for batch-level scaling.  

However, naive offloading can incur severe decoding delays and degrade overall throughput due to the limited access bandwidth of DRAM. Realizing efficient hierarchical memory management for DSAs requires addressing three key challenges that prior work has left open, as outlined in \S\ref{sec:introduction}:  
1) fragmented KV cache transfers that reduce effective DRAM bandwidth,  
2) lack of efficient runtime control for batch sizing, and  
3) the high HBM footprint of chunked prefill.

%% file: 3_design.tex
\section{The \name{} System}

In this section, we present \name{}, an LLM serving system with efficient hierarchical HBM-DRAM management designed for unlocking the parallel potential of DSAs. 

\subsection{System Overview}\label{subsec:overview}

\begin{figure}[!t]
	\centering
	\includegraphics[width=0.9\columnwidth]{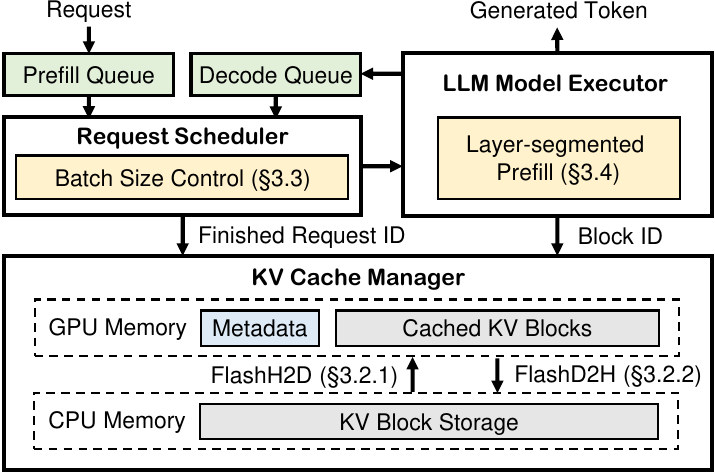}
	\caption{The system architecture for \name{}.}
	\label{fig:design:architecture}
    \vspace{-4mm}
\end{figure}

Figure~\ref{fig:design:architecture} illustrates the overall system architecture for \name{}, which comprises three key components: the request scheduler, the model executor, and the KV cache manager.

\squishlist

\item[\ding{182}] \textit{Request Scheduler.} The request scheduler determines which requests are executed in each iteration. Similar to existing LLM serving systems, it adopts dynamic batching techniques~\cite{orca} and schedules requests in a first-come-first-served (FCFS) manner. In addition, the scheduler in \name{} incorporates a working-set-aware batch size control strategy ($\S$~\ref{subsec:schedule}) to prevent GPU cache thrashing by avoiding excessive KV block loading. Additionally, the scheduler leverages the layer-segmented prefill technique ($\S$~\ref{subsec:layersegment}), which divides the layers of the prefill phase of a request into distinct segments. These segments are executed in separate batches, thereby reducing the memory footprint and limiting the runtime of each iteration. 

\item[\ding{183}] \textit{Model Executor.} The model executor performs the computation of model forwards. It replaces the standard attention with the sparse attention for decoding requests. It also tracks the prefill layers to execute in each batch and skips the other layers based on the layer-segmented prefill strategy ($\S$~\ref{subsec:layersegment}). During execution, the model executor communicates with the KV cache manager to transmit newly generated KV cache data and retrieve the \textit{metadata} of the relevant KV blocks. In addition, it sends the indices of the KV blocks required for the attention computation to the KV cache manager to trigger the KV cache loading from DRAM to HBM ($\S$~\ref{subsubsec:load}). The \textit{metadata} is used to estimate the criticality of KV blocks for each query token. By default, \name{} adopts the cuboid-mean method to construct the metadata for KV blocks due to its high accuracy~\cite{arkvale}. However, other metadata construction methods~\cite{infllm, quest} can be easily integrated into \name{}.

\item[\ding{184}] \textit{KV Cache Manager.} The KV cache manager maintains a hierarchical KV cache between HBM and DRAM. Both HBM and DRAM are organized into fixed-size blocks to mitigate memory fragmentation~\cite{vllm} and are managed independently per attention head. The KV cache manager receives newly generated KV caches from the model executor and saves them to the corresponding KV blocks in the DRAM ($\S$~\ref{subsubsec:save}). Once a KV block reaches its capacity, its associated metadata is created. The metadata is retained in HBM due to its small size and is utilized in every attention computation. The remaining HBM is used to cache frequently accessed KV blocks and we employ the least recently used (LRU) cache eviction policy, which leverages the cosine similarity between consecutive query tokens~\cite{infllm, tokenselect}. Specifically, query vectors of consecutive tokens exhibit high similarity, leading to the selection of similar KV blocks. 

\squishend

\subsection{Fragmentation-Aware KV Cache Transfer}\label{subsec:io}

\begin{figure}[!t]
	\centering
	\includegraphics[width=1.0\columnwidth]{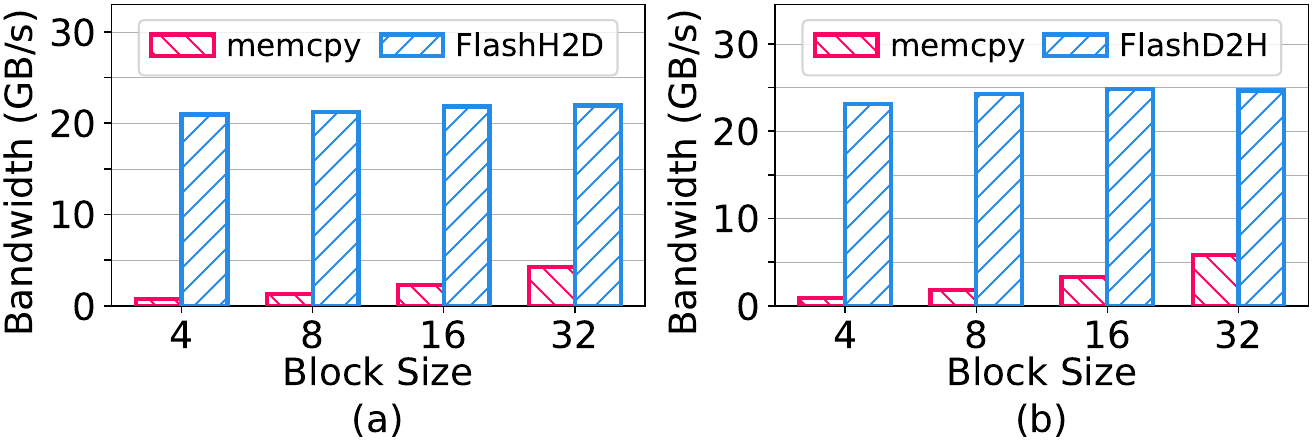}
	\caption{PCIe bandwidth of KV cache (a) loading with \texttt{memcpy} and FlashH2D, and (b) saving with \texttt{memcpy} and FlashD2H, under varying block sizes.}
    
	\label{fig:design:bandwidth}
    \vspace{-4mm}
\end{figure}

\begin{figure}[!t]
	\centering
	\includegraphics[width=0.9\columnwidth]{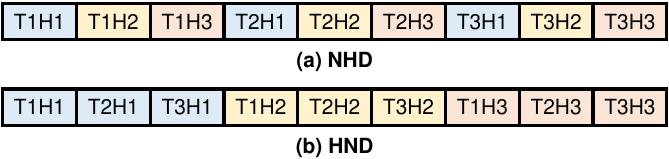}
	\caption{A comparison between the $(N,H,D)$ and $(H,N,D)$ KV block layouts. $N$, $H$, and $D$ denote the token, head, and hidden dimension, respectively.}
    
	\label{fig:design:layout}
    \vspace{-4mm}
\end{figure}

There are two layouts to organize the KV cache of each token in a KV block: $(N,H,D)$ and $(H,N,D)$. Figure~\ref{fig:design:layout} demonstrates this with an example of three tokens and three heads. In the $(N,H,D)$ layout, all KV heads of a token is stored contiguously. In contrast, the $(H,N,D)$ layout groups the KV cache for all tokens associated with a head together. Since DSAs select KV blocks at the head level, the $(H,N,D)$ layout is adopted for efficient block selection and memory access. 

With hierarchical HBM–DRAM storage for KV caches, both loading and saving KV blocks to DRAM can stall LLM computation. This bottleneck is further exacerbated by fragmented KV cache accesses, which occurs at the granularity of KV heads. To mitigate the overhead of KV cache transfers, \name{} adopts FlashD2H\&H2D, a fragmentation-aware transfer mechanism that tailors GPU-direct loading and CPU-assisted saving to their unique characteristics.


\subsubsection{FlashH2D: GPU-Direct Loading}\label{subsubsec:load}

\begin{figure}[!t]
	\centering
	\includegraphics[width=0.8\columnwidth]{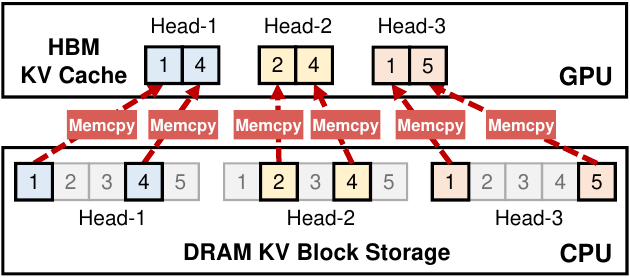}
	\caption{A demonstration of the fragmented KV cache block loading from DRAM to HBM using a \texttt{memcpy}-based method.}
	\label{fig:design:loading}
    \vspace{-3mm}
\end{figure}

Before attention computation in each model layer, DSAs dynamically select and load KV blocks from DRAM to HBM on a per-head basis. Figure~\ref{fig:design:loading} illustrates this process using the conventional \texttt{memcpy}-based approach. Because \texttt{memcpy} requires contiguous source and destination buffers, each KV block scattered in DRAM must be copied individually, resulting in many \texttt{memcpy} calls. This causes substantial function invocation overhead and poor PCIe bandwidth utilization, especially when the number of KV blocks is large.

To accelerate such fragmented KV cache loading, \name{} employs \emph{a GPU-direct loading strategy}, by leveraging unified virtual addressing (UVA) supported by modern GPUs. UVA allows GPU kernels to directly access the DRAM. Instead of issuing multiple \texttt{memcpy} calls, \name{} launches a single GPU kernel that loads all selected KV blocks from DRAM in parallel. The kernel assigns one thread block per KV block, adapting the number of threads dynamically to the actual number of selected KV blocks in each iteration. By fusing all KV block loads into one GPU kernel, GPU-direct loading minimizes invocation overhead and effectively increases PCIe bandwidth utilization. As shown in Figure~\ref{fig:design:bandwidth}a, FlashH2D consistently delivers PCIe bandwidth exceeding 20 GB/s across varying block sizes, significantly outperforming \texttt{memcpy}, whose bandwidth stays under 5 GB/s.

\subsubsection{FlashD2H: CPU-Assisted Saving}\label{subsubsec:save}

\begin{figure}[!t]
	\setlength{\abovecaptionskip}{0.1cm}
	\centering
	\includegraphics[width=0.75\columnwidth]{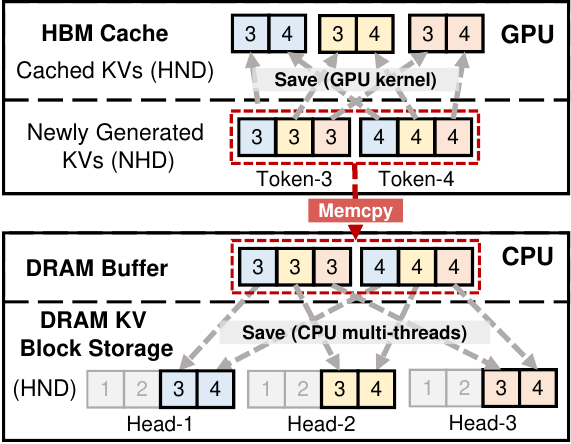}
	\caption{A demonstration of the execution workflow of the proposed CPU-assisted KV cache saving.}
	\label{fig:design:saving}
    \vspace{-3mm}
\end{figure}

At the beginning of each model layer, the hidden states from the previous layer are projected into a KV tensor of shape $(B, H, D)$, where $B$ denotes the total number of tokens in the current batch, $H$ is the number of attention heads, and $D$ is the head dimension. The resulting KV tensor is then saved into the free slots in the corresponding KV blocks in HBM. Once a KV block is full, it is asynchronously flushed into DRAM using a separate CUDA stream, overlapping saving with model computation. However, since the full KV blocks in HBM are scattered across non-continuous addresses,  saving suffers from the same fragmented data movement issue as loading. As a result, the saving latency can exceed the model computation latency. This problem is more prominent during prefill, when many KV caches are generated at the same time. One might consider using GPU kernels to accelerate saving as with loading. However, this approach consumes GPU resources and interferes with model computation, leading to prolonged execution time.

To address this issue, we propose \emph{CPU-assisted KV cache saving} in \name{}, as illustrated in Figure~\ref{fig:design:saving}. Our key observation is that the KV tensor generated in each iteration is continuous before saving to the KV blocks. Leveraging this property, \name{} decomposes saving into two steps: (1) the contiguous KV tensor is first copied into a DRAM buffer with a single \texttt{memcpy}, and (2) once the transfer completes, CPU threads redistributed the buffered data into the corresponding DRAM KV blocks. Crucially, the CPU-assisted saving method avoids consuming GPU computation resources, allowing saving to execute fully in parallel with model computation without any interference. As shown in Figure~\ref{fig:design:bandwidth}b, FlashD2H consistently delivers PCIe bandwidth exceeding 23 GB/s across varying block sizes, significantly outperforming \texttt{memcpy}, whose bandwidth stays under 6 GB/s.

\begin{figure}[!t]
	\centering
        \setlength{\abovecaptionskip}{0.1cm}
	\includegraphics[width=0.9\columnwidth]{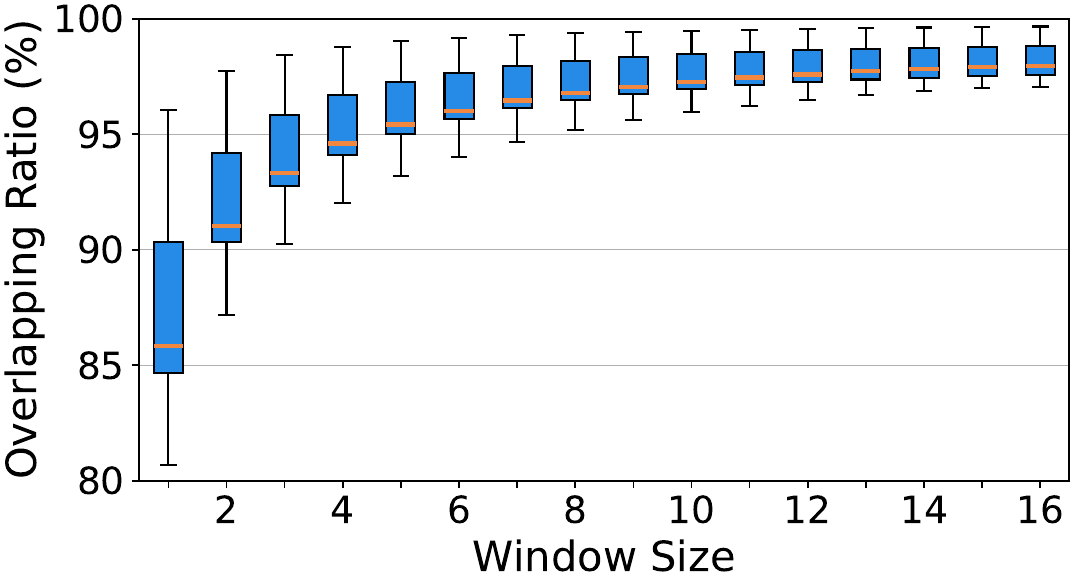}
	\caption{Average overlap ratios between the KV blocks accessed in preceding decoding steps and those selected in the current decoding step. We refer to the number of preceding decoding steps considered as window size.}
	\label{fig:design:workingset}
    \vspace{-2mm}
\end{figure}

\subsection{Working-Set-Aware Batch Size Control}\label{subsec:schedule}

To determine the optimal request batch size, \name{} employs \emph{a working-set-aware batch size control strategy} that dynamically adjusts batch sizes at runtime. The key idea is to estimate whether the working set—the total HBM capacity required by the KV cache of all running requests in the iteration—fits within HBM, thereby avoiding cache thrashing and excessive KV block loading. 
We first present how to estimate working set sizes for prefill and decoding requests, and then describe the scheduling workflow.


\begin{algorithm}[t]
	\caption{The scheduling algorithm.}
	\label{alg:schedulealgo}
	\DontPrintSemicolon
	\SetAlgoLined
	\LinesNumbered
	\SetNoFillComment
	\SetKwComment{Comment}{$\triangleright$\ }{}

        $\mathcal{R}_{max}$: The maximum number of requests per batch. \\
        $\mathcal{T}_{max}$: The maximum number of tokens per batch. \\
        $\mathcal{M}_{avl}$: The available HBM size. \\
        $\mathcal{S}$: The scheduler in existing LLM serving systems. \\
    
	\KwOut{The batch $\mathcal{B}_{res}$ for the next iteration.}
	\BlankLine
	
	$\mathcal{B}_{init}$
        $\leftarrow$ $\mathcal{S}.$\texttt{getBatch($\mathcal{R}_{max}$, $\mathcal{T}_{max}$)} \\

        $\mathcal{B}_{res}$ $\leftarrow$ $\{\}$ \\
        $\mathcal{M}_{used}$ $\leftarrow$ $0$ \\
        
        \For{\textnormal{request}$\ req$\ $\in$\ $\mathcal{B}_{init}$}{
            $\mathcal{M}_{req}$ $\leftarrow$ \texttt{estimateWS($req$)} \\
            \If{$\mathcal{M}_{used}$\ $+$\ $\mathcal{M}_{req}$\ $\le$\ $\mathcal{M}_{avl}$} {
               $\mathcal{B}_{res}.$\texttt{add($req$)} \\
               $\mathcal{M}_{used}$ $\leftarrow$ $\mathcal{M}_{used}$\ $+$\ $\mathcal{M}_{req}$
            } \Else {
               $\mathcal{S}.$\texttt{reset($req$)} \\
            }
        }

        \KwRet{$\mathcal{B}_{res}$}
\end{algorithm}

\stitle{Prefill working set}
The working set of a prefill request is the total HBM capacity required to store its KV cache during the current iteration. This value can be computed exactly, as the prefill process is deterministic. However, the working set size varies depending on the prefill strategy. For chunked prefill, the working set includes the KV cache from all preceding token chunks across all layers, since each chunk depends on previously generated KV caches. In contrast, for the layer-segmented prefill proposed in this paper ($\S$~\ref{subsec:layersegment}), the working set consists of only one layer of KV cache, since the KV caches of all previous layers are not required and can be evicted into DRAM once processed.

\stitle{Decoding working set}
Unlike prefill, the working set size of a decoding request cannot be directly calculated since the selected KV blocks at each decoding step vary dynamically. To address this, \name{} estimates the working set size by leveraging strong temporal locality: consecutive query tokens often select highly overlapping KV blocks~\cite{arkvale, infllm, tokenselect}.

We measure the average overlap ratios between KV blocks selected in the current decoding step and those selected in the preceding decoding steps (Figure~\ref{fig:design:workingset}). In the experiment, we vary the history window size (the number of preceding decoding steps considered). Experiments on LWM-7B~\cite{lwm} model across LongBench datasets~\cite{longbench} show consistently high overlaps. The overlap ratios increase sharply with the window size initially, but soon plateau. For example, expanding the window from 1 to 12 steps improves overlap by 10.68\%, while growing it further from 12 to 16 adds only 0.31\%. This suggests that it is sufficient to retain only a bounded history. Accordingly, \name{} tracks the KV blocks selected over the past $w$ decoding steps (with $w = 12$ by default) and regards their union as the decoding working set.

\begin{figure}[!t]
	\centering
	\includegraphics[width=1\columnwidth]{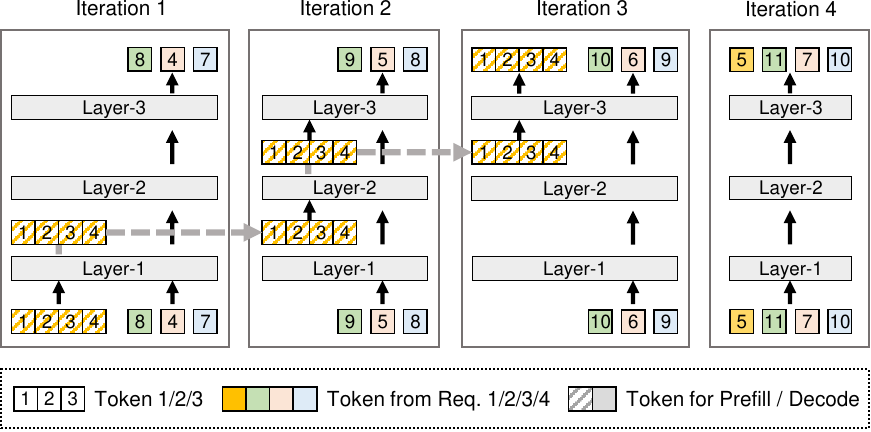}
	\caption{The workflow of layer-segmented prefill with a running example. (The LLM model consists of three layers. Each batch contains three decoding requests and one prefill request, with segment size set to 1.)}
	\label{fig:design:layer-segmented-prefilling}
     \vspace{-3mm}
\end{figure}

\stitle{Scheduling workflow}
We implement working-set-aware batch size control in \name{} by extending the scheduler of existing LLM serving systems, ensuring compatibility with prior designs. The scheduler operates under three input constraints for each iteration: (1) $\mathcal{R}_{max}$, which bounds the maximum number of requests in a batch; (2) $\mathcal{T}_{max}$, which limits the total number of tokens in a batch to control the computational workload, especially for prefill requests; and (3) $\mathcal{M}_{avl}$, which specifies the available GPU HBM cache capacity. The first two constraints are commonly used in existing systems. In contrast, \name{} introduces $\mathcal{M}_{avl}$, to ensure that the HBM usage of each request batch remains within cache limits, thereby preventing cache thrashing and HBM contention among requests.

To form a batch, \name{} first invokes the existing scheduling logic to construct an initial candidate batch that satisfies $\mathcal{R}_{max}$ and $\mathcal{T}_{max}$ constraints (Line~5). Then, \name{} estimates the working set size for each request in this initial batch, and adds the request to the current execution batch only if the total HBM usage within $\mathcal{M}_{avl}$ (Lines~8-12). If not, the request is rejected and its state is reset (Lines~13-14).

\subsection{Layer-Segmented Prefill}\label{subsec:layersegment}

To achieve both low HBM consumption and low TBT when used with DSAs, \name{} introduces \textit{layer-segmented prefill}, a new prefill mechanism designed for hybrid batching. The key observation is that LLMs are composed of multiple layers and the model forwarding is conducted layer by layer. Although the entire prefill of a long input prompt can be time-consuming, the execution of a single layer is relatively fast. This suggests that we can divide prefill into layer segments and process these segments in separate batches, bounding the runtime of each batch. Importantly, since the input prompt is not chunked, the KV blocks of each layer are accessed only once during prefill. The KV blocks of all finished layers can be immediately evicted after being saved to DRAM and the released HBM space can be reused for subsequent layers. This design bounds the HBM footprint of prefill to a single layer at any given time.

Figure~\ref{fig:design:layer-segmented-prefilling} demonstrates the workflow of the proposed layer-segmented prefill with a running example. The LLM model consists of three layers and each batch includes three decoding requests alongside one prefill request (the last iteration contains four decoding requests). With a segment size of 1, prefill is completed over three iterations (batches). Each iteration executes only one prefill layer alongside decoding requests and skips the remaining layers to ensure low batch running time. After executing each iteration except for the last one, activation states from the executed prefill layer are saved and used to resume prefill in the next iteration.

\stitle{Determining segment size}
The appropriate segment size for layer-segmented prefill depends on three factors: prompt length, TBT SLO, and system throughput. To achieve a low TBT, a small segment size is preferred, especially for long prompts. However, this slows down prefill and reduces throughput. To meet the varying demands in practice, \name{} provides a configurable parameter called \textit{maxInjectToken}, which limits the maximum number of prefill tokens injected into a batch. In practice, users can find the value for \textit{maxInjectToken} through profiling experiments. Specifically, users can set \textit{maxInjectToken} to a small initial value and gradually increase it until reaching the TBT limit, thus maximizing throughput under the given SLO.

\stitle{Combination with chunked prefill}
Layer-segmented prefill can be combined with chunked prefill for extremely long input prompts. If a single layer’s execution time already exceeds the TBT, we partition each layer into smaller chunks like chunked prefill, and process these chunks across multiple batches. This hybrid strategy provides fine-grained latency control while retaining the HBM efficiency benefits of layer-wise segmentation.

%% file: 5_evaluation.tex
\section{Performance Evaluation}
\subsection{Experimental Setup}
\stitlestart{Testbed}
Our experiments are conducted on a machine hosting an Nvidia A100 GPU with 40 GB HBM, an AMD EPYC 7J13 CPU, and 256 GB DRAM. The GPU is connected to the host via PCIe Gen 4, providing a bandwidth of 32 GB/s. 

\stitle{Models}
The experiments evaluate two popular long-context LLM models: LWM-7B~\cite{lwm} with a 1M context window and Llama3-8B~\cite{gradientlongcontextllama3} with a 262K context window. In particular, the LWM-7B model employs the same model architecture as Llama2-7B~\cite{llama2}. These two models cover two mainstream attention methods, namely, multi-head attention (MHA) and grouped query attention (GQA).

\stitle{Baselines}
\name{} is implemented based on vLLM~\cite{vllm}, which is a state-of-the-art LLM serving system that employs full KV cache attention without sparsity. We use vLLM as the first baseline for performance comparison. Next, we implement dynamic sparse attentions~\cite{arkvale} on vLLM as the second baseline, which is referred to as vLLM-S. In addition, we further enhance vLLM-S with KV cache offloading, which results in the third baseline called vLLM-SO.

\stitle{Workload}
We conduct experiments using multiple datasets from LongBench~\cite{longbench}, which covers various task types including \textit{Question Answering} (Qasper~\cite{qasper}, NarrativeQA~\cite{narrativeqa}, MultifieldQA~\cite{2wikimultihopqa}, Dureader~\cite{dureader}), \textit{Document Summarization} (GovReport~\cite{govreport}, QMSum~\cite{qmsum}, MultiNews~\cite{multinews}, VCSum~\cite{vcsum}), and \textit{Code Generation} (LCC~\cite{lcc}, RepoBench-P~\cite{repobench}). We combine the requests from all datasets into one trace to reflect real-world LLM serving scenarios where requests from different tasks are handled simultaneously. Following prior works~\cite{orca, sarathiserve}, we generate request arrival times based on a Poisson distribution with varying arrival rates, donated as request rate. To prevent vLLM from aborting requests with KV cache sizes exceeding HBM capacity, we limit the maximum prompt length of the requests. For LWM-7B and Llama3-8B, the maximum length is set to 32k and 128k, respectively.


\subsection{End-to-End Performance}


We evaluate the end-to-end performance of different schemes to demonstrate the efficiency of \name{}. For \name{}, vLLM-S, and vLLM-SO, the token budget of attention KV cache is set to 2,048 for LWM-7B and Llama3-8B, ensuring that the accuracy of sparse attention achieves $99\%$ of full attention, which is typical in production scenarios~\cite{AdaptiveNN, Apparate}. Due to page limit, we report the detailed accuracy of \name{} under various token budgets in supplemental materials. For vLLM, vLLM-S, and vLLM-SO, we use a token chunk size of 2,048 for chunked prefill. For a fair comparison, the layer-segmented prefill of \name{} processes the same number of tokens as the chunked prefill in each iteration by setting the \textit{maxInjectToken} as $B \cdot L$, where $B$ is the chunk size for chunked prefill and $L$ is the number of model layers.

\begin{figure}[!t]
    	\centering
	\includegraphics[width=1\columnwidth]{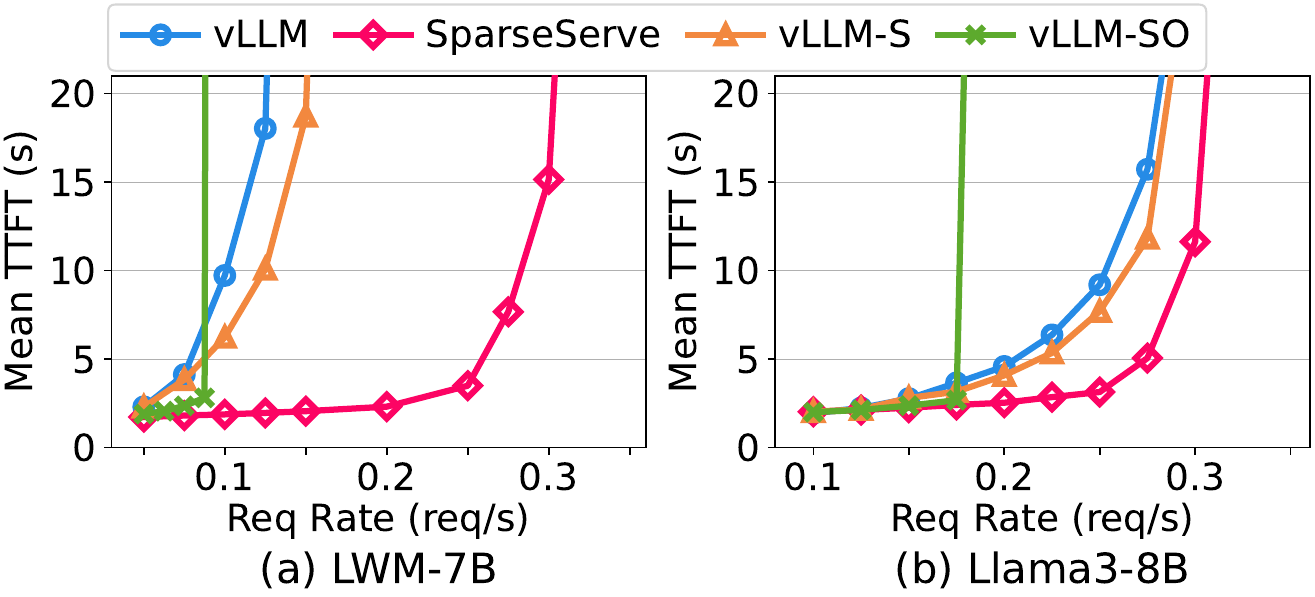}
	\caption{The mean TTFT of all systems under varying request rate on LongBench with LWM-7B and Llama3-8B.}
	\label{fig:eval:main_ttft}
	 \vspace{-2mm}
\end{figure}

\stitle{TTFT}
Figure~\ref{fig:eval:main_ttft} shows the mean TTFT of all systems under varying request rates. At low request rates, the TTFTs of all systems are similar. However, as the request rate increases, vLLM quickly exhausts HBM with its KV cache, blocking new requests and significantly prolonging queuing time. For example, at 0.125 req/s for LWM-7B, the TTFT of vLLM is 9.26$\times$ higher than that of \name{}. Sparse attention enables vLLM-S to reduce TTFT compared with vLLM due to faster decoding, while vLLM-SO further lowers TTFT at low rates by supporting larger batches via KV offloading. However, at high request rates, the TTFT of vLLM-SO becomes worse than vLLM and vLLM-S due to excessive KV block loading latency. In contrast, \name{} consistently achieves the lowest TTFT across request rates by combining sparse attention with its system-level optimizations.


\begin{figure}[!t]
	\centering
	\includegraphics[width=1\columnwidth]{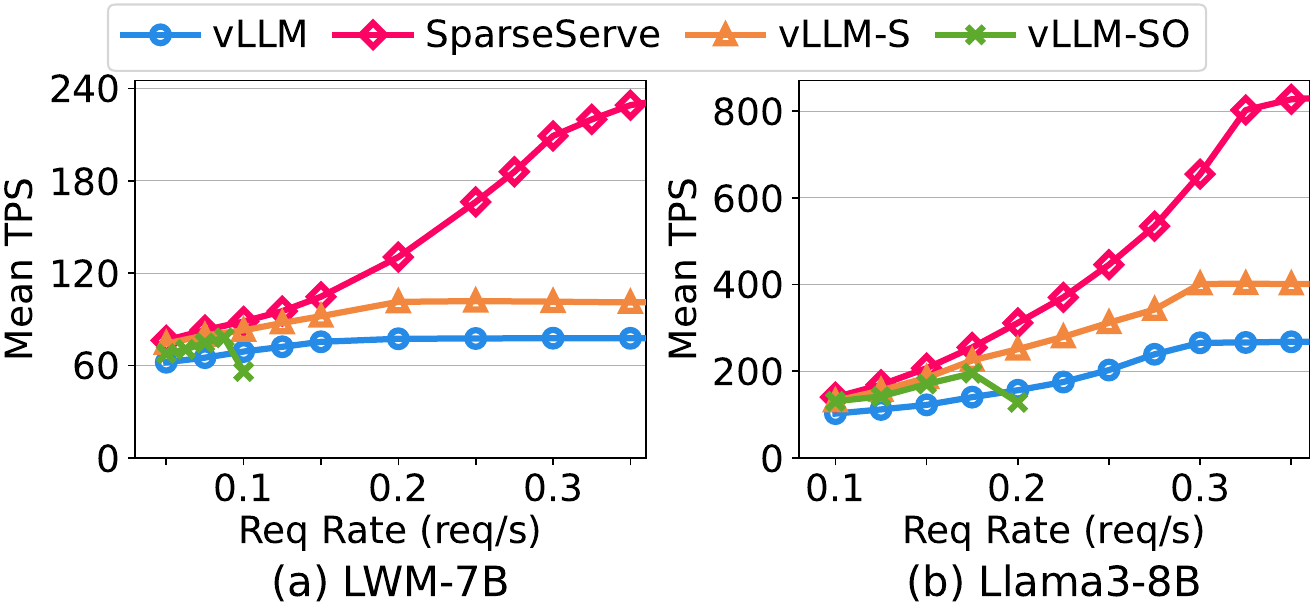}
	\caption{The mean token generation throughput of all systems under varying request rates on LongBench with LWM-7B and Llama3-8B models.}
	\label{fig:eval:main_throughput}
    \vspace{-2mm}
\end{figure}

\stitlestart{Token generation throughput}
Figure~\ref{fig:eval:main_throughput} reports the token generation throughput. Due to the excessive running time of vLLM-SO under high request rates, we cap the maximum request rates of vLLM-SO at 0.1 and 0.2 RPS for LWM-7B and Llama3-8B, respectively. All systems achieve similar throughput at low request rates. As the request rate increases, the throughput of vLLM and vLLM-S reaches a plateau due to the small batches limited by the HBM. By applying sparse attention, vLLM-S achieves higher throughput than vLLM due to lower decoding latency. Although vLLM-SO enables larger batch sizes by offloading KV cache to DRAM, its throughput is worse than vLLM-S due to heavy KV cache loading overhead. In contrast, \name{} consistently delivers the highest throughput compared with all baselines. Specifically, compared with vLLM, \name{} achieves throughput improvements of up to 2.93$\times$ and 3.14$\times$ for LWM-7B and Llama3-8B, respectively. Against vLLM-S, \name{} achieves up to 2.23$\times$ and 2.03$\times$ improvements, and against vLLM-SO, up to 2.96$\times$ and 4.24$\times$, respectively.

\begin{figure}[!t]
	\centering
	\includegraphics[width=1\columnwidth]{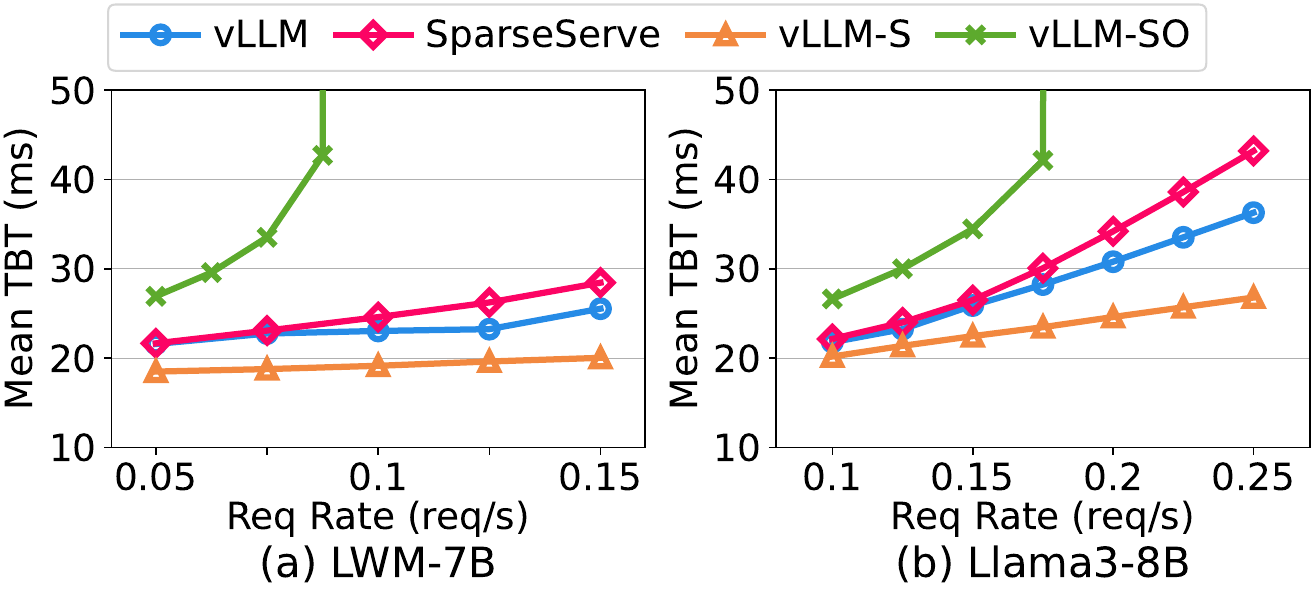}
	\caption{The mean TBT of all systems under varying request rates on LongBench with LWM-7B and Llama3-8B.}
	\label{fig:eval:main_tbt}
	\vspace{-2mm}
\end{figure}

\stitle{TBT} Figure~\ref{fig:eval:main_tbt} shows the mean TBT of all systems under varying request rates. To prevent the excessive request queuing of vLLM, we cap its maximum request rates at 0.15 and 0.25 RPS for LWM-7B and Llama3-8B, respectively. Compared with vLLM, vLLM-S reduces attention computation time and maintains the same average batch sizes, thus achieving lower TBT. Due to the larger average batch sizes and the heavy KV cache loading overhead, vLLM-SO achieves the highest TBT among all systems, which is consistent with the results in Figures~\ref{fig:eval:main_ttft} and~\ref{fig:eval:main_throughput}. Thanks to the proposed system designs, \name{} effectively reduces the KV block loading overhead and achieves slightly higher TBT than vLLM, which is within 20\% for both LWM-7B and Llama3-8B. We argue that this minor TBT degradation is a reasonable trade-off, as \name{} achieves significantly lower TTFT and higher token generation throughput compared to vLLM.

\begin{figure}[!t]
	\centering
	\includegraphics[width=1\columnwidth]{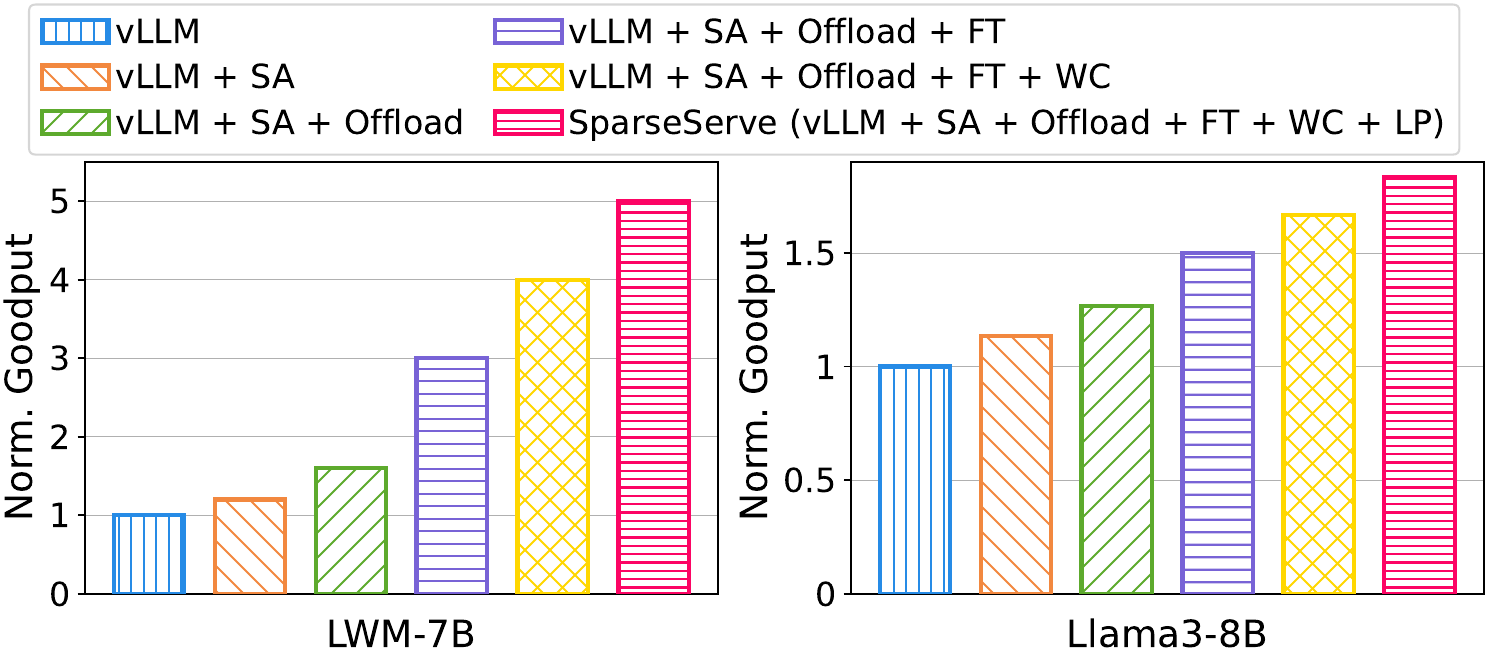}
	\caption{The maximum request throughput under SLO requirement of the designs in \name{} under varying input request rates on LongBench with LWM-7B and Llama3-8B models (SA: sparse attention; Offload: KV cache offloading; FT: fragmentation-aware KV cache transfer; WC: working-set-aware batch size control; LP: layer-segmented prefill).}
	\label{fig:eval:main_goodput}
    \vspace{-3mm}
\end{figure}

\stitle{Goodput}
Finally, we analyze the incremental impact of \name{}’s design components on goodput, defined as the maximum sustainable request throughput under SLOs, as shown in Figure~\ref{fig:eval:main_goodput}.
Following prior work~\cite{splitwise, sarathiserve}, we define the SLOs for the P99 TBT as 25$\times$ the execution time of a decoding iteration. 
In addition, we ensure the sustainability of the maximum input request load by imposing a threshold on request queuing delay. Specifically, the mean scheduling delay for input requests is limited to 2 seconds to prevent excessive queuing, as suggested in~\cite{sarathiserve}.

Starting from vLLM, adding sparse attention (\textbf{vLLM+SA}) improves goodput by 1.20$\times$ (LWM-7B) and 1.13$\times$ (Llama3-8B) by reducing decoding latency. Offloading (\textbf{vLLM+SA+Offload}) further boosts goodput by 1.33$\times$ and 1.12$\times$ by lowering HBM usage and enabling larger batches. Fragmentation-aware transfer (\textbf{vLLM+SA+Offload+FT}) brings larger gains (1.88$\times$ and 1.19$\times$) by accelerating fragmented KV transfers and tolerating higher cache miss rates. Working-set-aware batch control (\textbf{vLLM+SA+Offload+FT+WC}) contributes an additional 1.33$\times$ and 1.11$\times$ improvement by preventing GPU cache thrashing. Finally, layer-segmented prefill (\textbf{vLLM+SA+Offload+FT+WC+LP}) completes \name{}, improving goodput by another 1.25$\times$ and 1.10$\times$ by reducing prefill memory demands and lowering queuing delays for long prompts. Collectively, these optimizations enable \name{} to improve goodput by up to 5.00$\times$ on LWM-7B and 1.83$\times$ on Llama3-8B compared with vLLM.

\subsection{Ablation Studies}
We conduct ablation studies with LWM-7B on LongBench to isolate the effects of the three proposed designs in \name{}: fragmentation-aware KV cache transfer, working-set-aware batch size control, and layer-segmented prefill.

\begin{figure}[!t]
	\setlength{\abovecaptionskip}{0.05cm}
	\centering
	\includegraphics[width=\columnwidth]{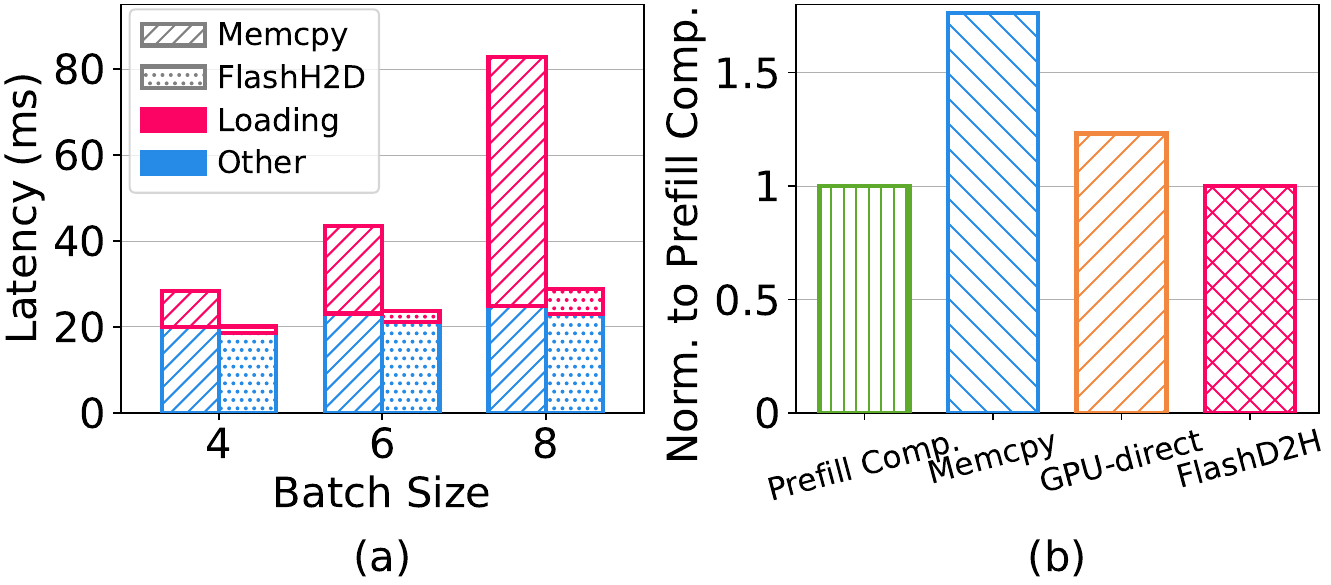}
\caption{(a) The mean batch latency and the mean KV cache loading latency of \texttt{memcpy}-base and FlashH2D with varying running batch sizes. (b) The mean prefill latencies of \texttt{memcpy}-based, GPU-direct, and FlashD2H normalized to the standalone prefill computation time.}
	\label{fig:eval:loading}
	\vspace{-2mm}
\end{figure}

\subsubsection{Fragmentation-Aware KV Cache Transfer}\hfill


\stitle{FlashH2D: GPU-direct KV cache loading}
Figure~\ref{fig:eval:loading}a shows the mean batch latency and KV cache loading latency between \texttt{memcpy}-based loading and FlashH2D. We observe that the KV cache loading time accounts for a substantial portion of the total batch latency, particularly at larger batch sizes. For example, when the batch size is 8, the KV cache loading accounts for 69.94\% of the total batch latency. In contrast, FlashH2D effectively eliminates this bottleneck, reducing KV cache loading latency by up to 9.97$\times$ compared to the \texttt{memcpy}-based baseline.

\stitle{FlashD2H: CPU-assisted KV cache saving}
Figure~\ref{fig:eval:loading}b shows the mean prefill latency, normalized to the standalone prefill computation time, for three KV cache saving methods: \texttt{memcpy}-based saving, GPU-direct saving, and FlashD2H. Prefill is chosen as the evaluation phase because it generates a large volume of new KV cache blocks, making the saving overhead more prominent. We execute the model computation and KV cache saving with different CUDA streams to overlap their executions. We observe that the mean prefill latency with the \texttt{memcpy}-based method is 1.76$\times$ longer than the prefill computation time. This overhead results from fragmented KV block saving via \texttt{memcpy}, which cannot be fully hidden by computation. For the GPU-direct saving method, the prefill latency is 1.28$\times$ longer than the computation time. The reason is that GPU-direct saving consumes GPU resources, introducing contention that prolongs the model computation phase. In contrast, the prefill latency with FlashD2H is the same as the prefill computation time, indicating that saving newly generated KV cache to DRAM can be fully overlapped with the prefill computation, thus introducing no overhead.

\subsubsection{Working-Set-Aware Batch Size Control}\hfill

\begin{figure}[!t]
        \setlength{\abovecaptionskip}{0.05cm}
	\centering
	\includegraphics[width=\columnwidth]{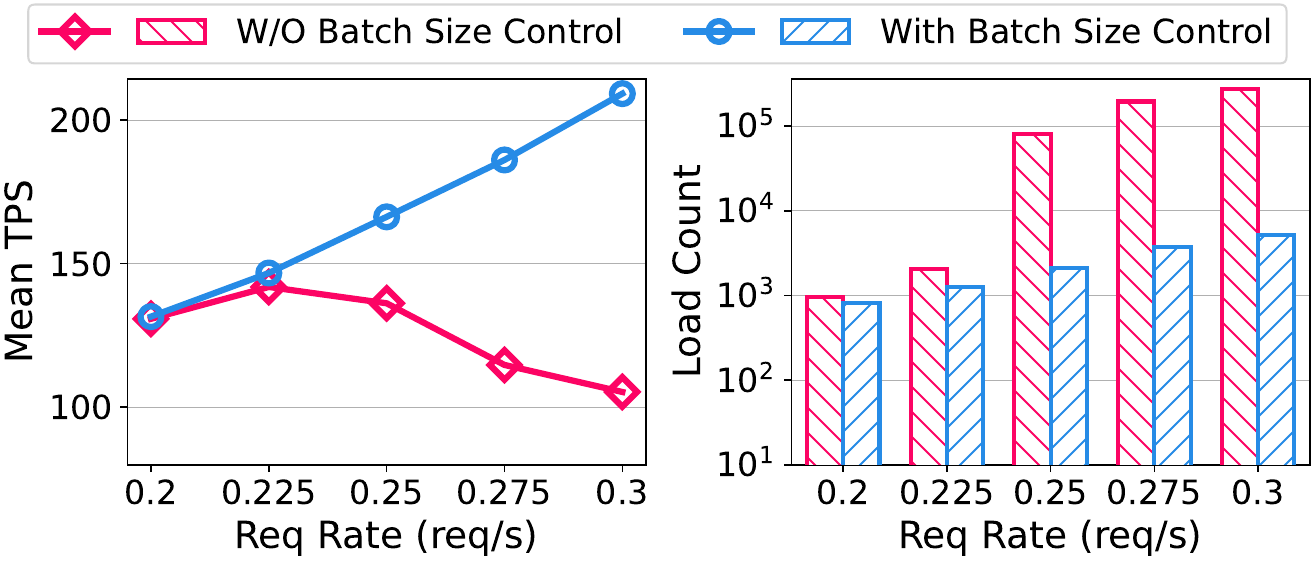}
	\caption{The throughput as well as the mean KV block loading numbers with and without the working-set-aware batch size control under varying request rates.}
	\label{fig:eval:schedule}
    \vspace{-2mm}
\end{figure}

\noindent We evaluate the proposed working-set-aware batch size control by measuring token generation throughput and the mean number of KV block loads per iteration, as shown in Figure~\ref{fig:eval:schedule}. At low request rates, the token generation throughput is similar with and without batch size control. This is because the batch sizes are small and the GPU cache is sufficient to hold the working sets of all scheduled requests, resulting in few KV block loads. However, as the request rate increases, the throughput of the baseline (without batch size control) starts to decrease. For example, when the request rate increases from 0.25 to 0.3 RPS, the throughput drops by 29.52\% due to a sharp increase in KV block loads, as shown on the right side of Figure~\ref{fig:eval:schedule}. In contrast, working-set-aware control mitigates cache contention, cutting KV block loads by 52.78$\times$ at 0.3 RPS. Consequently, throughput continues to increase steadily with higher request rates.

\subsubsection{Layer-Segmented vs. Chunked Prefill}\hfill

\begin{figure}[!t]
        \setlength{\abovecaptionskip}{0.05cm}
	\centering
	\includegraphics[width=\columnwidth]{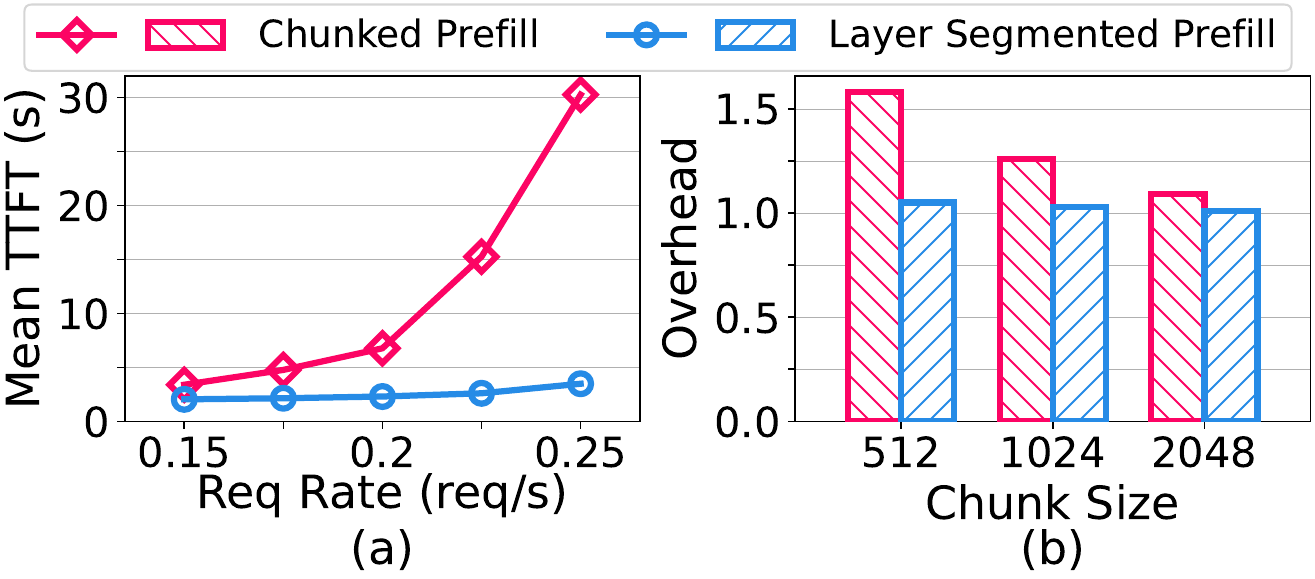}
	\caption{(a) The mean TTFT of chunked prefill and layer-segmented prefill under varying request rates. (b) The overhead of chunked prefill and layer-segmented prefill in attention computation normalized to the cost of plain prefill with varying token chunk sizes.}
	\label{fig:eval:lpcpttft}
     \vspace{-2mm}
\end{figure}


\stitle{TTFT reduction}
Figure~\ref{fig:eval:lpcpttft}a shows the mean TTFT with the proposed layer-segmented prefill and chunked prefill under varying request rates. The chunk size is set to 2,048 for chunked prefill. For a fair comparison, we configure the layer-segmented prefill to process the same number of tokens as the chunked prefill in each iteration. We observe that when the request rate is low, the TTFT of layer-segmented prefill is similar to that of chunked prefill. However, by increasing the request rate, the number of ongoing decoding requests increases and increases the HBM utilization, which starts blocking prefill execution due to HBM shortage. In contrast, layer-segment prefill reduces the HBM requirement during prefilling, lowering queuing time and TTFT. It reduces the mean TTFT by up to 8.68$\times$, compared to chunked prefill.

\stitle{Prefill computation overhead}
Figure~\ref{fig:eval:lpcpttft}b shows the overhead of chunked prefill and the layer-segmented prefill compared to plain prefill regarding the attention time during prefilling. We observe that chunked prefill incurs high overhead when the chunk size is small. Specifically, with a chunk size of 512, chunked prefill slows down the prefill attention by 1.51$\times$. This is because chunked prefill necessitates the repeated loading of the KV cache of all preceding chunks to process the latest chunk. In contrast, layer-segmented prefill exhibits performance nearly identical to that of plain prefill, thereby minimizing the overhead.

%% file: 6_related_work.tex
\section{Related Work}\label{sec:related-work}
\stitlestart{Static sparse attention}
Several KV cache eviction algorithms, e.g., H2O~\cite{h2o}, StreamingLLM~\cite{streamingllm}, SnapKV~\cite{snapkv}, FastGen~\cite{fastgen}, and Scissorhands~\cite{scissorhands}, have been proposed to retain only the KV cache of important tokens while discarding others to save HBM. However, since the importance of tokens changes during the decoding process, discarded tokens may become crucial for future computation~\cite{quest}, resulting in potential accuracy loss. 

\stitle{Dynamic sparse attention} DSAs, such as ArkVale~\cite{arkvale}, InfLLM~\cite{infllm}, Quest~\cite{quest}, mitigate the issue of static sparse attention by dynamically selecting a small portion of the critical KV cache for attention computation for each query token while retaining all KV cache. While existing DSAs predominantly focus on enhancing the accuracy of important KV cache identification, \name{} is the first work to consider their deployment efficiency in practical LLM serving systems, achieving both high accuracy and efficiency through system designs tailored for DSAs.

\stitle{Token-level sparse attention} Recent works, such as InfiniGen~\cite{infinigen}, TokenSelect~\cite{tokenselect}, RetrievalAttention~\cite{retrievalattention}, and MagicPig~\cite{magicpig}, perform KV cache selection at the granularity of tokens. Although token-level selection can identify important tokens more accurately in theory, it is overly granular and incurs significant runtime overhead~\cite{arkvale}. In contrast, block-level selection achieves a balance between accuracy and performance overhead, as a result of which, this paper focuses on block-level DSA approaches.


\stitle{Sparse attention for prefill and training} There are also works~\cite{Gems, mInference, SeerAttention} that apply sparse attention to accelerate the prefill phase and training of LLMs. GemFilter~\cite{Gems} prunes unimportant tokens using attention matrices from early layers to reduce the computational load in subsequent layers. Minference~\cite{mInference} recognizes three general patterns of sparse attention in long-context LLMs and provides optimized CUDA kernels for each pattern. SeerAttention~\cite{SeerAttention} extends Minference by replacing fixed patterns with a learnable approach. Native sparse attention (NSA)~\cite{Yuan2025NativeSA} introduced by DeepSeek is the first to perform sparse attention during training. These methods are orthogonal to our \name{} and can be combined with it to further enhance the efficiency of end-to-end LLM processing.

\stitle{Inference parameter offloading}
DeepSpeed Inference~\cite{deepspeed-inference, ZeRO-Infinity} offloads model parameters to DRAM and fetches them on demand. Lina~\cite{lina} leverages sparse activation in mixture-of-experts (MoE) models to offload cold experts to DRAM. PowerInfer~\cite{PowerInfer} utilizes the sparsity in FFN computation to offload inactive weights to DRAM, saving HBM and computational resources. FlexGen~\cite{flexgen} offloads both model parameters and KV cache to DRAM, targeting offline processing. In contrast, \name{} exploits KV cache offloading in online LLM serving by utilizing the sparsity in KV cache.


%% file: 7_conclusion.tex
\section{Conclusion}
This paper presents \name{}, an efficient long-context LLM serving system that unlocks the parallel potential of DSAs through efficient hierarchical HBM–DRAM management. To achieve efficient and scalable DSA deployment, \name{} incorporates three core techniques, including fragmentation-aware KV cache transfer, working-set-aware batch size control, and layer-segmented prefill. Extensive experimental results demonstrate that \name{} reduces the mean TTFT by up to 9.26$\times$ and increases the token generation throughput by up to 3.14$\times$ compared to state-of-the-art LLM serving systems. 

%% file: appendices.tex
\appendix
\section{Appendices}\label{sec:appendices}

\subsection{Model Accuracy}

\begin{table*}[!hb]
    \footnotesize
    \setlength{\abovecaptionskip}{0.15cm}
    \caption{Model accuracy with varying token budgets.}
    \label{tab:model-comparison}
    \centering
    \setlength{\tabcolsep}{4pt}
    \fontsize{9}{11}\selectfont
    \begin{tabular}{c|ccccc|ccccc}
        \toprule
        & \multicolumn{5}{c}{\textbf{LWM-Text-Chat-1M}} & \multicolumn{5}{c}{\textbf{Llama-3-8B-262k}} \\
        \cmidrule(lr){2-6} \cmidrule(lr){7-11}
        \textbf{Dataset} & \textbf{Full} & \textbf{0.5k} & \textbf{1k} & \textbf{1.5k} & \textbf{2k} & 
                          \textbf{Full} & \textbf{0.5k} & \textbf{1k} & \textbf{1.5k} & \textbf{2k} \\
        \midrule
        HotpotQA        & 21.93 & \cellcolor{green!30}22.22 & \cellcolor{green!30}22.54 & \cellcolor{green!30}22.16 & \cellcolor{green!30}22.22 &
                          17.79 & \cellcolor{green!30}17.99 & \cellcolor{green!30}17.70 & \cellcolor{green!30}17.86 & \cellcolor{green!30}17.83 \\
        2WikiMultihopQA & 18.01 & \cellcolor{green!30}18.22 & \cellcolor{red!30}17.67 & \cellcolor{green!30}18.13 & \cellcolor{green!30}18.47 &
                          16.90 & \cellcolor{red!30}16.51 & \cellcolor{red!30}16.46 & \cellcolor{green!30}16.70 & \cellcolor{green!30}16.77 \\
        MuSiQue         & 10.36 & \cellcolor{green!30}10.78 & \cellcolor{green!30}10.37 & \cellcolor{green!30}10.65 & \cellcolor{green!30}10.25 &
                           9.52 & \cellcolor{red!30}9.17 & \cellcolor{red!30}9.17 & \cellcolor{green!30}9.42 & \cellcolor{green!30}9.83 \\
        DuReader        & 25.68 & \cellcolor{green!30}25.71 & \cellcolor{green!30}27.02 & \cellcolor{green!30}26.29 & \cellcolor{green!30}26.03 &
                          27.47 & \cellcolor{green!30}27.25 & \cellcolor{green!30}27.79 & \cellcolor{green!30}27.76 & \cellcolor{green!30}27.47 \\
        MultiFieldQA-en & 43.05 & \cellcolor{green!30}43.38 & \cellcolor{green!30}43.77 & \cellcolor{green!30}42.77 & \cellcolor{green!30}42.51 &
                          40.27 & \cellcolor{green!30}40.24 & \cellcolor{green!30}41.19 & \cellcolor{green!30}40.56 & \cellcolor{green!30}40.31 \\
        NarrativeQA     & 13.50 & \cellcolor{red!30}13.15  & \cellcolor{green!30}13.79 & \cellcolor{green!30}13.64 & \cellcolor{green!30}13.68 &
                          16.99 & \cellcolor{green!30}17.14 & \cellcolor{red!30}16.60 & \cellcolor{green!30}16.76 & \cellcolor{green!30}16.95 \\
        Qasper          & 24.08 & \cellcolor{green!30}24.12 & \cellcolor{green!30}25.06 & \cellcolor{green!30}24.62 & \cellcolor{green!30}24.49 &
                          26.21 & \cellcolor{red!30}25.54 & \cellcolor{red!30}25.59 & \cellcolor{green!30}25.93 & \cellcolor{green!30}26.23 \\
        GovReport       & 27.88 & \cellcolor{red!30}26.36 & \cellcolor{red!30}27.07 & \cellcolor{red!30}27.35 & \cellcolor{green!30}27.48 &
                          33.61 & \cellcolor{red!30}30.01 & \cellcolor{red!30}31.95 & \cellcolor{red!30}32.63 & \cellcolor{green!30}33.19 \\
        QMSum           & 24.83 & \cellcolor{red!30}24.30 & \cellcolor{green!30}24.63 & \cellcolor{green!30}24.85 & \cellcolor{green!30}24.95 &
                          25.51 & \cellcolor{red!30}25.09 & \cellcolor{green!30}25.80 & \cellcolor{green!30}25.69 & \cellcolor{green!30}25.54 \\
        MultiNews       & 24.38 & \cellcolor{red!30}23.62 & \cellcolor{red!30}23.98 & \cellcolor{red!30}24.00 & \cellcolor{green!30}24.32 &
                          27.87 & \cellcolor{red!30}27.34 & \cellcolor{green!30}27.74 & \cellcolor{green!30}27.90 & \cellcolor{green!30}27.75 \\
        VCSUM           &  9.52 & \cellcolor{green!30}11.09 & \cellcolor{green!30}10.57 & \cellcolor{green!30}10.66 & \cellcolor{green!30}10.45 &
                          14.54 & \cellcolor{red!30}14.15 & \cellcolor{red!30}13.89 & \cellcolor{red!30}13.98 & \cellcolor{green!30}14.36 \\
        TriviaQA        & 61.87 & \cellcolor{red!30}60.57 & \cellcolor{red!30}60.76 & \cellcolor{green!30}62.01 & \cellcolor{green!30}61.65 &
                          85.83 & \cellcolor{green!30}86.46 & \cellcolor{green!30}86.41 & \cellcolor{green!30}86.38 & \cellcolor{green!30}86.44 \\
        SAMSum          & 39.91 & \cellcolor{green!30}39.56 & \cellcolor{green!30}39.77 & \cellcolor{green!30}39.93 & \cellcolor{green!30}39.68 &
                          41.62 & \cellcolor{red!30}40.48 & \cellcolor{green!30}41.17 & \cellcolor{green!30}41.29 & \cellcolor{green!30}41.02 \\
        LSHT            & 24.00 & \cellcolor{red!30}22.00 & \cellcolor{red!30}23.00 & \cellcolor{red!30}23.50 & \cellcolor{green!30}24.00 &
                          43.50 & \cellcolor{red!30}35.50 & \cellcolor{red!30}39.50 & \cellcolor{red!30}42.00 & \cellcolor{green!30}43.00 \\
        LCC             & 40.47 & \cellcolor{red!30}39.15 & \cellcolor{green!30}40.20 & \cellcolor{green!30}40.14 & \cellcolor{green!30}40.21 &
                          51.04 & \cellcolor{green!30}50.38 & \cellcolor{green!30}51.14 & \cellcolor{green!30}51.03 & \cellcolor{green!30}51.02 \\
        RepoBench-P     & 42.78 & \cellcolor{red!30}41.56 & \cellcolor{green!30}42.75 & \cellcolor{green!30}42.65 & \cellcolor{green!30}42.84 &
                          44.79 & \cellcolor{green!30}44.46 & \cellcolor{green!30}45.48 & \cellcolor{green!30}45.04 & \cellcolor{green!30}44.75 \\
        \bottomrule
    \end{tabular}
\end{table*}

We evaluate model accuracy on the LongBench~\cite{longbench} datasets under varying token budgets for attention computation, the results are shown in Table~\ref{tab:model-comparison}. The experiments are conducted using the state-of-the-art DSA ArkVale~\cite{arkvale}. We observe that both LWM-7B~\cite{lwm} and Llama3-8B~\cite{gradientlongcontextllama3} retain $99\%$ of the accuracy achieved with full attention across all datasets when the token budget is set to 2048. Based on this observation, we adopt a token budget of 2048 for sparse attention in our paper.

